\documentclass[journal=nalefd,manuscript=letter,layout=traditional]{achemso}
\usepackage{amsmath}
\usepackage{amssymb}
\usepackage{graphicx}
\usepackage{siunitx}
\usepackage{inputenc}
\usepackage{hyperref}

\graphicspath{{}{Figures/}}

\newcommand{\Vg}{V_\mathrm{g}}
\newcommand{\Ug}{U_\mathrm{g}}
\newcommand{\Ui}{U_\mathrm{i}}
\newcommand{\ddz}{\partial_z}

\title{Massive and topological surface states in tensile strained HgTe}

\author{David M.\ Mahler}
\author{Valentin L.\ M\"uller}
\author{Cornelius Thienel}
\author{Jonas Wiedenmann}
\author{Wouter Beugeling}
\author{Hartmut Buhmann}
\author{Laurens W.\ Molenkamp}
\affiliation{Institute for Topological Insulators and Physikalisches Institut, Experimentelle Physik III, Universit\"at W\"urzburg,
Am Hubland, 97074 W\"urzburg, Germany}


\begin{document}

\begin{abstract}
Magneto-transport measurements on gated high mobility heterostructures containing a 60 nm layer of tensile strained HgTe, a three-dimensional topological insulator, show well-developed Hall quantization from surface states both in the n- as well as in the p-type regime. While the n-type behavior is due to transport in the topological surface state of the material, we find from 8-orbital k$\cdot$p calculations that the p-type transport results from massive Volkov-Pankratov states. Their formation prevents the Dirac point and thus the p-conducting topological surface state from being accessible in transport experiments. This interpretation is supported by low-field magneto-transport experiments demonstrating the coexistence of n-conducting topological surface states and p-conducting Volkov-Pankratov states at the relevant gate voltages.
\end{abstract}

\maketitle

The notion that band inversion in a narrow-gap material can lead to Dirac-type surface states goes back to the 1980's \cite{Volkov1985, Kusmartsev1985} and while it was realized very early on that these states might originate from topology \cite{Fradkin1986} it took two more decades---and a detour involving the quantum spin Hall effect \cite{KaneMele2005PRL95-14,BernevigEA2006,Konig2007}---before it was realized that the surface states of topological insulators are indeed the gapless Dirac states discussed in Refs.~\cite{Volkov1985, Kusmartsev1985}.
While the gapless states of Ref.~\cite{Volkov1985} have been intensively studied in the last decade, showing clear signatures in ARPES
\cite{Chen2009,Xia2009} and in transport experiments in the quantum Hall regime \cite{Brune2011, Brune2014}, much less attention has been attracted by the massive surface states that were also first predicted in that paper. These massive Volkov-Pankratov states are the gapped solutions of the Schr\"odinger equation at an interface where the band inversion takes place smoothly over a finite width. They are pulled from the bulk to the surface in presence of a sufficiently large electric field perpendicular to the surface.
The states are thus topologically trivial.

The actual occurrence of such massive Volkov-Pankratov states in a topological insulator was inferred from high frequency compressibility experiments on strained bulk HgTe-based heterostructures \cite{Inhofer2017}, where a sudden drop in the diffusion constant at high gate voltage was interpreted as resulting
from scattering of the topological surface state with a massive Volkov-Pankratov state, generated by the large electric field at the surface of the HgTe layer.
While this indirect experimental signature is consistent with the presence of a massive Volkov-Pankratov state, direct evidence in the form of transport signatures has been elusive as the experiments were not designed to pin down the microscopics of the actual massive states.

In this paper, we show that it is possible to create p-type massive Volkov-Pankratov states at relatively small electric fields, and that they lead to a readily observable quantum Hall effect in (Hg,Cd)Te heterostructures containing a strained bulk HgTe layer that offer a much improved mobility (up to and above $\mu_\text{e}=\SI{500e3}{\cm^{2}/Vs}$ for electrons) as compared to Refs.~\cite{Brune2011, Brune2014, Inhofer2017}. At the same time, these samples show a clean n-type quantum Hall effect from the topological surface states at the relevant gate voltages.
In the p regime, a large density of low-mobility holes paired with a small density of high-mobility electrons directly points at the presence of Volkov-Pankratov states coexisting with the topological surface state.
Pinning of the Fermi level to the Volkov-Pankratov states lifts it up for increasingly negative gate voltages. This mechanism prevents the p-side of the topological surface states from being experimentally accessible.

Our (Hg,Cd)Te heterostructures have been grown by molecular beam epitaxy on CdTe substrates.  The lattice mismatch of 0.3\% between substrate and the HgTe layer
provides the required tensile strain to open a gap in the HgTe band structure, turning it into a high quality topological insulator, virtually without any bulk doping \cite{Brune2011, Brune2014}. The active region, a \SI{60}{\nm} thick HgTe layer, is sandwiched between a \SI{100}{\nm} buffer and \SI{5}{\nm} cap layer of Hg$_{0.3}$Cd$_{0.7}$Te.

\begin{figure*}
\centering
\includegraphics[width=420pt]{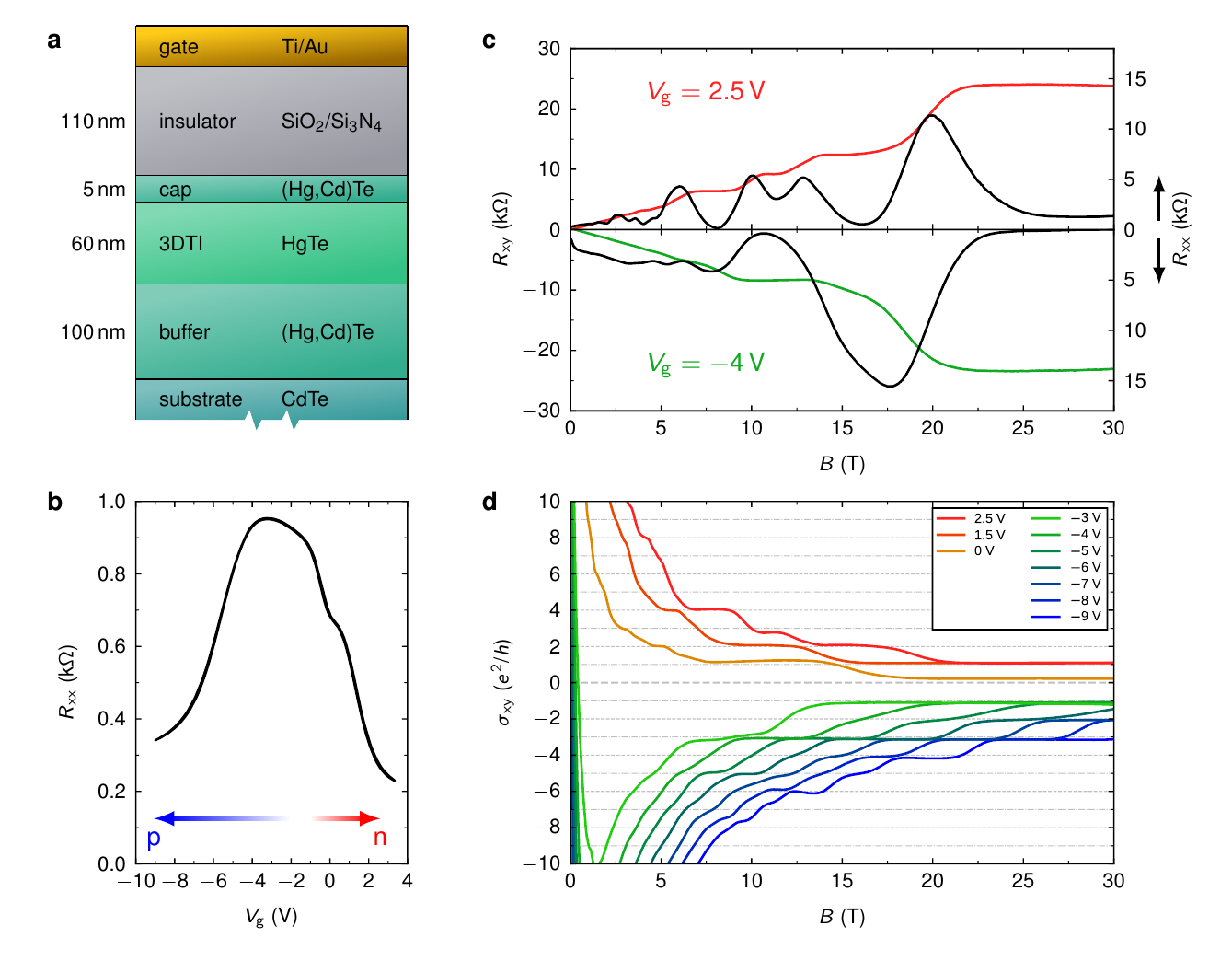}
\caption{
\textbf{Magneto-transport measurements for sample 1 at $T\sim\SI{500}{mK}$.}
\textbf{a} The device structure, with layer thickness indicated on the left-hand side.
\textbf{b} Gate sweep at zero field: The longitudinal resistance $R_\text{xx}$ as function of gate voltage $\Vg$. The arrows indicate the n and p regimes.
\textbf{c} The longitudinal $R_\text{xx}$ (black) and Hall resistance $R_\text{xy}$ (colored) as function of magnetic field $B$ for n density ($\Vg=\SI{2.5}{V}$) and p density ($\Vg=\SI{-4}{V}$).
\textbf{d} Hall conductivity $\sigma_\text{xy} = - R_\text{xy} /((R_\text{xx} / A )^2 + R_\text{xy}^2   )$ in units of $e^2/h$ for a more complete set of gate voltages $\Vg$ from $\SI{-9.0}{V}$ to $\SI{2.5}{V}$.
}
\label{fig:HighBSweeps}
\end{figure*}

The samples are lithographically shaped into Hall bar structures of $\SI{200}{\um}$ width and $\SI{600}{\um}$ length between the voltage probes. The Hall bars are covered with a top gate stack consisting of 11 alternating \SI{10}{\nm}-thick $\rm SiO_2$ and $\rm Si_3N_4$ layers serving as dielectric, a \SI{5}{\nm}/\SI{100}{\nm} thick $\rm Ti$/$\rm Au$ layer forming the gate electrode. The layer stack is displayed in Fig.~\ref{fig:HighBSweeps}a.
We present data for two samples (sample 1 and sample 2) fabricated from the same wafer.
Transport measurements are done with the standard low-frequency lock-in technique in a ${}^3{\rm He}$ cryostat (sample 1) and a ${}^3{\rm He}/{}^4{\rm He}$ dilution refrigerator (sample 2), at estimated electron temperatures of $\SI{500}{\milli\K}$ and $\SI{150}{\milli\K}$, respectively.
A magnetic field is applied perpendicular to the HgTe layer plane.

Figures~\ref{fig:HighBSweeps}b and c show the longitudinal, $R_\text{xx}$, and transverse, $R_\text{xy}$, resistance of sample~1 for gate voltages between $\Vg=\SI{-9.0}{V}$ and $\SI{2.5}{V}$, covering a total density range between $n_{tot}=\SI{-18.6e11}{\cm^{-2}}$ (p-regime) and $n_{tot}=\SI{8.0e11}{\cm^{-2}}$ (densities extracted from the linear part of the low magnetic field Hall resistance).
The plateaus of $R_\text{xy}$ and minima of $R_\text{xx}$ are clear indicators of Hall quantization in both the n- and p-conducting regimes.
In Fig.~\ref{fig:HighBSweeps}d, we plot the Hall conductivity $\sigma_{xy} = - R_\text{xy} / ( (R_\text{xx} / A  )^2 + R_\text{xy}^2)$ of the sample for several intermediate gate voltages, where $A$ is the geometric aspect ratio of the device (length divided by width). The interplay of different carrier systems is apparent from the observation that the sequence of Hall plateaus exhibits missing steps for given gate voltages, which reappear when the density is slightly changed. This observation was reported in Ref.~\cite{Brune2014} for n-type carriers and connected to different densities in top and bottom surface states.
Whilst Refs.~\cite{Brune2011, Brune2014} only reported quantum Hall data on n-type carriers, the enhanced mobility of the capped structures in the present study also allow access to quantum Hall states in the p-regime:
Here, for example, as we evolve from the $\Vg=\SI{-3.0}{V}$ trace (green line in Fig. ~\ref{fig:HighBSweeps}d) toward more negative gate voltage some of the missing p-type even plateaus ($\sigma_{xy}=-2e^2/h, -4e^2/h$) reappear.

To clarify the origin of the contributing carrier systems we compare our experiments with k$\cdot$p calculations using a basis of 8 orbitals \cite{Novik2005,Brune2014}.
We calculate dispersions and eigenstates in a layered geometry, confined and discretized in the growth direction $z$ and infinite in the $(x, y)$ directions.
We identify the localization of the wave functions from the expectation values $\langle z \rangle$, as to distinguish between bulk and surface states. In Fig.~\ref{fig:kdotpplots}, we present band structures for four different potential strengths $\Ui$. We use red and blue color to highlight states localized near the top or bottom surface, respectively.

\begin{figure*}
\centering
\includegraphics[width=468pt]{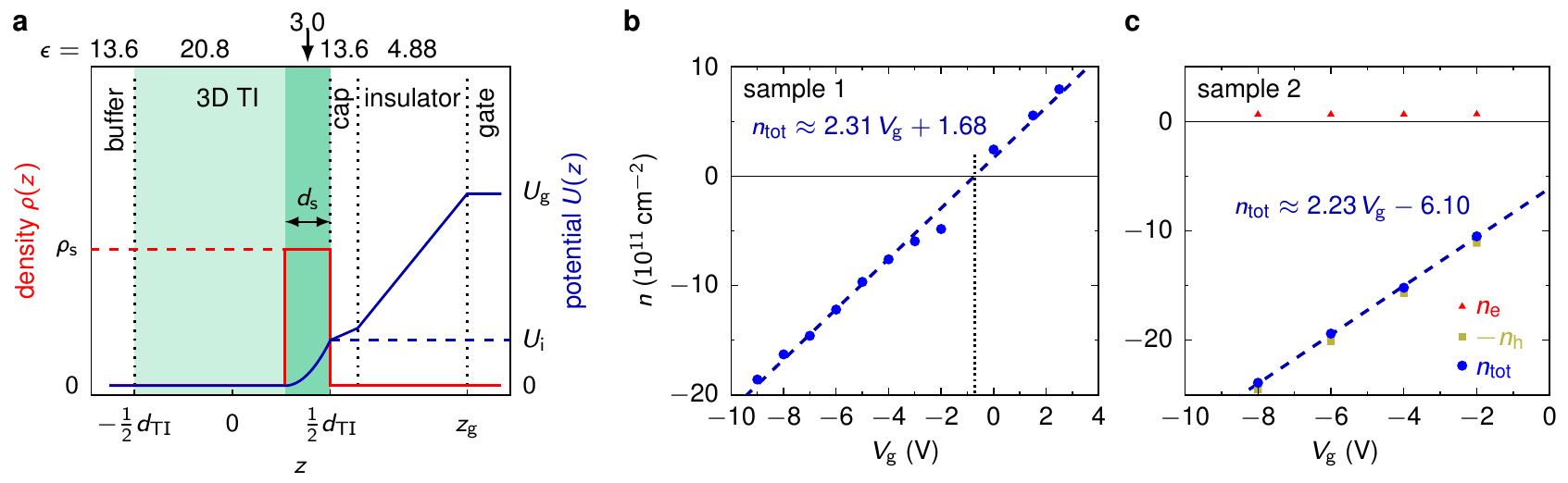}
\caption{
\textbf{Electrostatic effect of the gate voltage.}
\textbf{a} Charge density $\rho(z)$ and potential profile $U(z)$ as function of coordinate $z$ in growth direction. The value $z=0$ is the centre of the HgTe layer with thickness $d_\text{TI}=\SI{60}{nm}$; $z_\mathrm{g}$ is the $z$ coordinate of interface between insulator and gate. The shaded area is the 3D TI, with the surface-state region of thickness $d_\text{s}=\SI{8}{nm}$ indicated by darker shading. At the top, we indicate the values of the dielectric constant $\epsilon$.
\textbf{b} Total density $n_\text{tot} = n_\text{e}-n_\text{h}$ as function of the gate voltage $\Vg$ for sample 1 (data points). The dashed line indicates a linear fit, $n_\text{tot}\approx 2.31 \Vg+1.68$, with $n_\text{tot}$ in units of $10^{11}\,\text{cm}^{-2}$ and $\Vg$ in V. The vertical dotted line is the estimated charge neutrality point at $\Vg\approx\SI{-0.7}{V}$, defined by the intercept of the fit with $n_\text{tot}=0$.
\textbf{c} Total density $n_\text{tot}$, electron density $n_\text{e}$, and (negative) hole density $-n_\text{h}$ for sample 2.
}
\label{fig:Electrostatics}
\end{figure*}

\begin{figure}
\centering
\includegraphics[width=240pt]{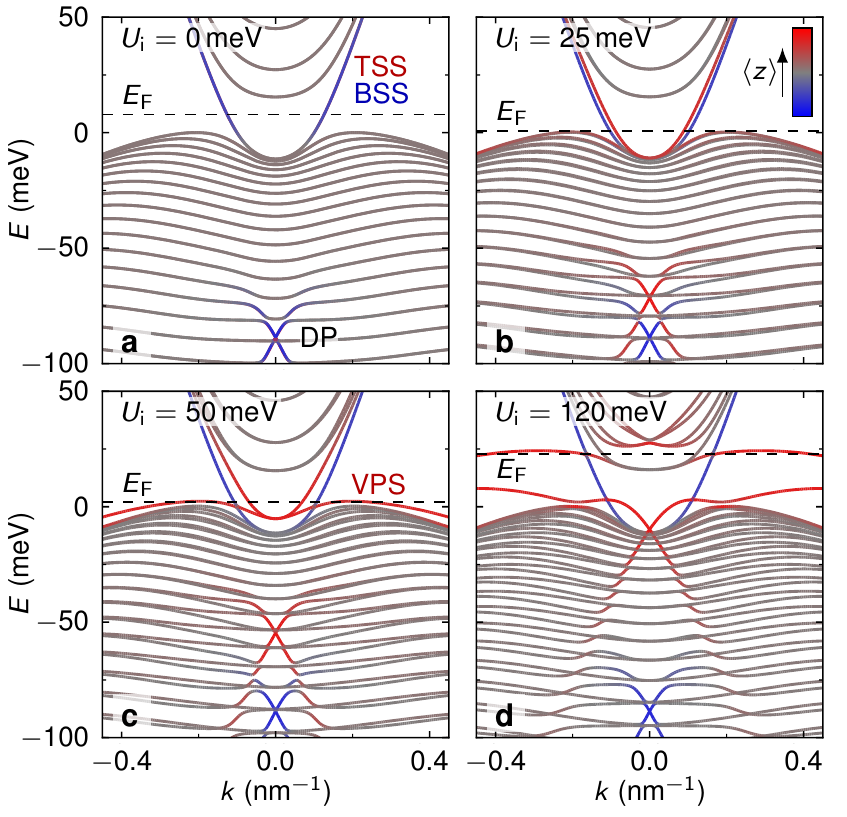}  
\caption{
\textbf{Band structures from $8$-orbital k$\cdot$p calculations for a \SI{60}{nm} thick tensilely strained HgTe layer.}
\textbf{a}  Band structure without electrostatic potential, we indicate the Dirac point (DP), top and bottom surface states (TSS and BSS, respectively). All states are doubly degenerate. The dashed line indicates the Fermi energy.
\textbf{b}, \textbf{c},  \textbf{d} Band structures under application of an electrostatic potential (potential strength $\Ui$ is indicated). In \textbf{c}, we label the Volkov-Pankratov state (VPS).
The color code [legend in \textbf{b}] indicates the wave function location (expectation value $\langle z \rangle$: top surface (red line), bottom surface (blue line), or gray elsewhere.
}
\label{fig:kdotpplots}
\end{figure}

We model the effect of the gate voltage by considering finite carrier densities near the top surface. The bulk being free of carriers is a natural assumption given the virtual absence of doping inside the material. For simplicity, we assume the charge density $\rho(z)$ to be uniform in a region of thickness $d_\text{s}=\SI{8}{nm}$ emanating from the top surface (see Fig.~\ref{fig:Electrostatics}a, red line).
The energy potential $U(z)$, related to electric potential $V(z)$ by $U(z)=-eV(z)$ (where $e$ is the electron charge), is solved from the Poisson equation,
\begin{equation}
  \ddz\bigl(\epsilon(z)\ddz U(z)\bigr) = \frac{e}{\epsilon_0}\rho(z),
\end{equation}
where $\ddz$ is the derivative in the $z$ direction and $\epsilon_0$ the vacuum permittivity; see Supporting Information for further details. The potential $U(z)$ has a quadratic functional dependence in this region and reaches $\Ui$ at the top interface, illustrated in Fig.~\ref{fig:Electrostatics}a (blue curve). The value at the gate is $\Ug\equiv U(z_\mathrm{g})=-e\Vg$.

In line with Ref.~\cite{Brune2014}, we have assumed the dielectric constant $\epsilon$ for the top surface state to be $\epsilon_\mathrm{s}= 3.0$, different from the bulk value $\epsilon_\mathrm{HgTe}=20.8$ for HgTe \cite{Baars1972}, as to model the strong screening of the top surface state. The response $\Delta \Ui$ of the potential to a gate voltage change $\Delta\Vg$ scales approximately as (see Supporting Information) 
\begin{equation}\label{eq:potentialgate}
   \Delta \Ui/\Delta\Vg \sim -\epsilon_\mathrm{s}^{-1},
\end{equation}
so that a lower dielectric constant means a stronger response. With the bottom surface kept at ground potential, we obtain a gate action $\Delta n / \Delta \Vg = \SI{2.4e11}{\cm^{-2}V^{-1}}$, consistent with the experimental value $\SI{2.31e11}{\cm^{-2}V^{-1}}$ extracted from the measured density as function of gate voltage, see Fig.~\ref{fig:Electrostatics}b. For the sake of illustration, we take into consideration the top surface only.

Importantly, the strength of the potential also determines the position of the Dirac point of the top surface state \cite{Brune2014}.
For $\Ui=0$, Fig.~\ref{fig:kdotpplots}a, we find the Dirac point is at \SI{-90}{meV} below the maximum of the valence band.
This location results \cite{Brune2014} from the band inversion of the $\Gamma_8$
and $\Gamma_6$ bands in $T_d$-symmetric unstrained bulk HgTe, which amounts to about \SI{300}{meV} \cite{Laurenti1990}. Lattice strain opens up a band gap (approximately \SI{20}{meV} at the $\Gamma$-point) between the heavy and light
$\Gamma_8$ subbands \cite{LiuLeung1975,Brune2011}. (Here, by definition, $E=0$ is defined as the band edge of $\Gamma_8$ for bulk HgTe.)
The valence band maximum is not at the $\Gamma$-point, but rather occurs at finite momentum (so that the gap is indirect), due to the combined effects of band inversion and hybridization \cite{Brune2014,Shamim2020SciAdv}. This feature, referred to as \emph{camel back}, results in a van Hove singularity in the density of states \cite{Ortner2002}.
At $\Ui=0$, the Fermi energy is approximately \SI{10}{meV} above the camel back, corresponding to an n-type density of $n\approx\SI{2.5e11}{cm^{-2}}$.

Upon application of a (negative) gate voltage, the Fermi energy decreases and reaches the level of the bulk valence band at about $\Ui \approx 25$ meV, while the surface state conductance remains n-type. The van Hove singularity (density of states $\approx \SI{4e11}{cm^{-2}\,meV^{-1}}$) at the camel back now pins the Fermi level (Fig.~\ref{fig:kdotpplots}b).
Because the electrostatic potential is asymmetric, the dispersions of the topological surface state on top (red) and bottom (blue) surface now become different, resulting in an imbalance in their occupation.
Despite the large density of bulk valence band states at the camel back, these do not contribute significantly to transport, due to their very low mobility. Indeed, plateaus of $\sigma_{xy}$ (Fig.~\ref{fig:HighBSweeps}d) can be resolved clearly, but do not align precisely with the quantized values.

\begin{figure}
\centering
\includegraphics[width=360pt]{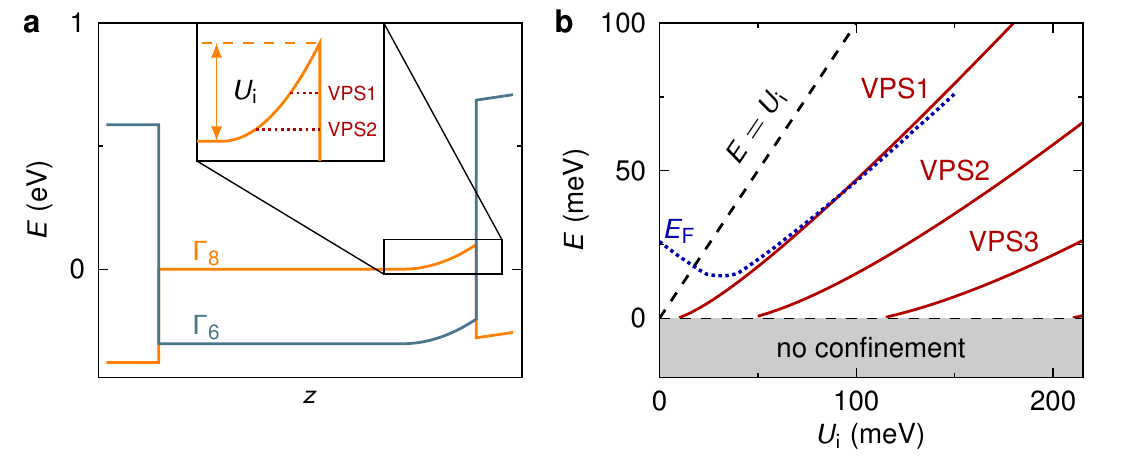}  
\caption{
\textbf{Confinement at the 3DTI top interface due to an electrostatic potential.}
\textbf{a}~Band edges of $\Gamma_6$ and $\Gamma_8$ shifted by the potential $U(z)$.
In the inset, we show a magnification of the $\Gamma_8$ band edge near the top interface of the 3D TI. The red dotted lines indicate confined hole states---these are the Volkov-Pankratov states.
\textbf{b} Energies of the states confined by the $\Gamma_8$ band edge as function of the potential strength $\Ui$, calculated within a simplified model with a single quadratic band. Here, the first Volkov-Pankratov state (VPS1) appears at a finite potential strength $\Ui\approx10\,\mathrm{meV}$. As $\Ui$ increases, more confined states (VPS2, VPS3, etc.) appear. This potential does not host confined states with $E<0\,\mathrm{meV}$. We sketch the Fermi energy (blue dotted curve), illustrating the pinning to the first Volkov-Pankratov state.
}
\label{fig:confinement}
\end{figure}

Upon further increasing the potential to $\Ui = 50$ meV (Fig.~\ref{fig:kdotpplots}c), we observe that a new state emerges from the valence band, localized at the top (red) surface.  The localization at the top surface is due to the confinement imposed by the gate potential, as illustrated in the inset of Fig.~\ref{fig:confinement}a).
(A similar mechanism has been explored in Ref.~\cite{LuGoerbig2020}.)
The emerging state can be identified as the first of the series of massive (trivial) surface states predicted by Volkov and Pankratov \cite{Volkov1985}.
The sudden emergence of this localized state at a finite value of $\Ui$ is due to the confinement energy of this state in the band bending due to the applied potential. 
In Fig.~\ref{fig:confinement}b, we show how the energies of the confined states evolve as function of the potential strength $\Ui$, within a simplified single-band model.
Since the dispersion of the emerging state is rather flat, similarly shaped as the (bulk) camel back, the high density of states $\approx \SI{2e11}{cm^{-2}\,meV^{-1}}$ will now in turn dominantly pin the Fermi level, illustrated by the dotted curve in Fig.~\ref{fig:confinement}b.
From this situation onwards any p-type quantum Hall effect observed in the magneto-transport experiment has to be attributed to this massive Volkov-Pankratov state at the top surface of the device.

Note that the position of the Fermi level implies that while the Volkov-Pankratov state is p-type and contains the majority of the carriers, at the same time the topological surface states contain a small amount of (high mobility) electrons, mostly in the bottom surface state.
For still larger electrostatic potentials (Fig.~\ref{fig:kdotpplots}d, $\Ui = 120$ meV), the Dirac point of the topological surface state enters into the band gap. However, the pinning of the Fermi level to the first emerging massive Volkov-Pankratov state prevents the observation of the p-conducting side of the topological surface states in a transport experiment. This is the main finding of this work: the experimentally observable p-type surface state transport, including the quantum Hall effect, of strained HgTe is entirely due to a massive Volkov-Pankratov state.

\begin{figure}
\centering
\includegraphics[width=240pt]{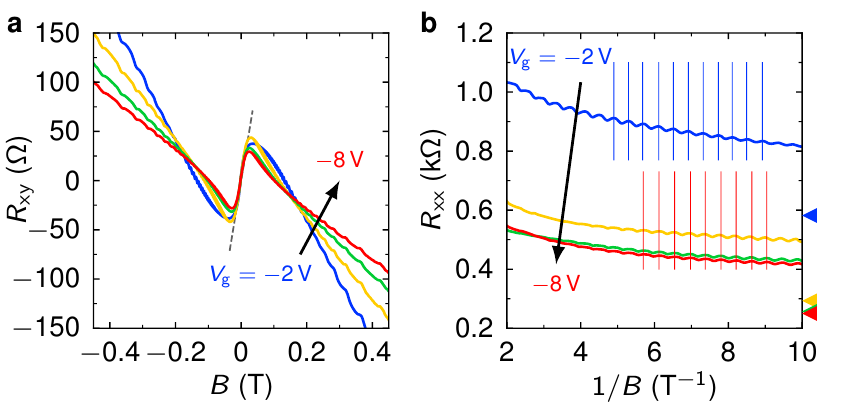}  
\caption{%
\textbf{Magneto-transport on fully strained HgTe layer at $T\sim\SI{150}{mK}$, sample 2.}
\textbf{a} $R_\text{xy}$ in low magnetic fields at different gate voltages (relative to the charge neutrality point, $\Vg= \SI{-2}{V},  \SI{-4}{V},  \SI{-6}{V},  \SI{-8}{V}$). Dashed lines indicate the low-magnetic field slope fitted to $R_\text{xy}$.
\textbf{b} $R_\text{xx}$ at the same gate voltages as shown in \textbf{a} as function of $1/B$. The markers on the right-hand side indicate the values of $R_\text{xx}$ at $B=\SI{0}{T}$. The vertical lines are aligned with the local maxima of the curves at $\Vg=\SI{-2}{V}$ and $\SI{-8}{V}$, as an aid to extract $f_{B^{-1}}$.
}
\label{fig:lowBsweeps}
\end{figure}

Apart from the quantum Hall behavior, the magneto-transport data of our devices in the relevant gate voltage ranges
contain additional experimental evidence for the role of massive surface states.
A first indication of the change of carrier origin when entering from the n- into the p-dominated transport regime is a strong reduction of carrier mobility. From the data presented in the Supporting Information, a reduction of the carrier mobility from $\mu_\text{tot}=\SI{90e3}{\cm^{2}/Vs}$ at $\Vg=\SI{2.5}{V}$ to $\mu_\text{tot}=\SI{32e3}{\cm^{2}/Vs}$ at $\Vg=\SI{-9.0}{V}$ is observed. In case of a p-conducting topological surface state one would expect the mobility remains roughly the same (cf.\ Fig.~\ref{fig:kdotpplots}d). Obviously, the large effective mass of the massive Volkov-Pankratov state readily explains our observation.

Additionally, with the high density of states of the massive Volkov-Pankratov state pinning the Fermi level, one expects that an increasingly negative gate voltage should affect only the carrier density in the Volkov-Pankratov state itself, while leaving the density of the topological surface states that coexist at the same energy roughly constant. In order to investigate this assumption we examine the low magnetic field Hall resistance in Fig.~\ref{fig:lowBsweeps}.
As shown in Fig.~\ref{fig:lowBsweeps}a, the Hall resistance of sample 2 at an estimated electron temperature of $\SI{150}{mK}$ in the magnetic field range between $B= -0.4$ and $0.4$\,T displays two distinct slopes: a steep n-type slope around zero $B$-field and a sharp turnover into a p-type slope for higher magnetic fields. While the p-type slope changes with gate voltage, the n-type slope exhibits no notable change within the investigated range, in agreement with expectations---the Volkov-Pankratov state accommodates all
extra carriers induced by changing the gate voltage.

The longitudinal resistance $R_\text{xx}$ reveals a series of oscillations, see Fig.~\ref{fig:lowBsweeps}b, where we have highlighted the periodicity in $1/B$. [The $1/B$ oscillations are also faintly visible in $R_\text{xy}$. Since resistivity and conductivity are related by matrix inversion, the oscillations originating in the longitudinal conductivity $\sigma_\text{xx}$ also affect the transversal resistance $R_\text{xy}$.]
We may attribute these oscillations to the high mobility electrons of one of the topological surface states. From this Figure it also is evident that the oscillation period does not change significantly within this applied gate-voltage range (indicated by the vertical lines as a guide for the eye). This directly implies that the responsible carriers are in the topological surface state at the bottom of the device structure---the field from the top gate is effectively screened by the mobile carriers at the top surface \cite{Brune2014}.
From the $1/B$-periodicity $f_{B^{-1}}$ of the oscillations, one can readily extract the related carrier density $n_\text{e} = \frac{e}{h} f_{B^{-1}}$, which varies only very slightly between $n_\text{e}=\SI{0.60e11}{\cm^{-2}}$ at $\Vg=\SI{-2}{V}$ and $n_\text{e}=\SI{0.57e11}{\cm^{-2}}$ at $\Vg=\SI{-8}{V}$.
The total density $n_\text{tot}=n_\text{e}-n_\text{h}$ is extracted from the linear Hall slope at magnetic fields $B>\SI{0.1}{T}$ (see Fig.~\ref{fig:Electrostatics}c and Supporting Information), from which we also obtain the hole densities $n_\text{h}$.

The mobilities $\mu_{\text{e},\text{h}}$ of the electron and hole conductance channels are extracted from the value of $R_\text{xx}$ and the slope of $R_\text{xy}$ at zero field, using a two-carrier Drude model (we again assume that transport in the topological surface state at the top surface can be neglected),
\begin{equation}\label{eq:twocarrlowB}
\begin{aligned}
\left. R_\text{xx}\right\vert_{B=0}
  &= \frac{A}{e \left(\mu_\text{e} |n_\text{e}| + \mu_\text{h} |n_\text{h}|\right)},\\
\left. \frac{d R_\text{xy}}{dB}\right\vert_{B=0}
  &= \frac{\mu_\text{e}^2 n_\text{e}-\mu_\text{h}^2 n_\text{h}}{e \left( \mu_\text{e} |n_\text{e}| + \mu_\text{h} |n_\text{h}|  \right)^2},
\end{aligned}
\end{equation}
where 
the already known densities $n_{\text{e},\text{h}}$ are substituted. From Eq.~\ref{eq:twocarrlowB} we solve the carrier mobilities $\mu_\text{e}=\SI{556e3}{\cm^{2}/Vs}$ for electrons and $\mu_\text{h}=\SI{18e3}{\cm^{2}/Vs}$ for holes, in qualitative agreement with the difference in carrier effective mass inferred from the band structure calculations.

While the quantized Hall conductivity of Fig.~\ref{fig:HighBSweeps}d points to  effectively two-dimensional surface states, the extracted density and mobility values indicate that charge transport for strongly negative gate voltages is dominated by low-mobility holes, paired with a lower density of highly mobile electrons. These properties are consistent with massive and massless surface states, respectively.
The combination of the transport observations thus provides direct experimental evidence of the coexistence of massive Volkov-Pankratov states with a massless topological surface state at the bottom surface.

In conclusion, we have shown convincing evidence of the interplay between the massless topological surface state and massive Volkov-Pankratov states in high
resolution magneto-transport experiments on the strained HgTe. While the specifics of massive state formation will vary between topological insulator materials, our experiments illustrate the care that must be taken to ensure that any observation of carrier transport at the surface of a topological material is indeed due
to a topological surface state, especially when high gate voltages need to be used to induce the surface state transport.


\begin{acknowledgement}
We thank E.~M. Hankiewicz, J.~B\"{o}ttcher, S.~Shamim, and L.-X.~Wang for discussions and C.~Ames, P.~Leubner, and L.~Lunczer for assistance with sample fabrication.
We acknowledge financial support from the Deutsche Forschungsgemeinschaft
(DFG, German Research Foundation) in the Leibniz Program and in the projects SFB
1170 (Project ID 258499086) and SPP 1666 (Project ID 220179758), from the EU ERC-AdG
program (Project 4-TOPS),
from the W\"urzburg-Dresden Cluster of Excellence on Complexity and Topology in Quantum Matter (EXC 2147, project ID 39085490),
and from the Free State of Bavaria (Elitenetzwerk
Bayern IDK `Topologische Isolatoren' and the Institute for Topological Insulators).
Part of the measurements were performed at the High Field Magnet Laboratory (HFML), Nijmegen, Netherlands.
\end{acknowledgement}


\begin{suppinfo}
Modelling of the electrostatics, Properties of sample 1, Properties of sample 2
\end{suppinfo}


\section{Author contributions}
H.B. and L.W.M. planned the experiments.
C.T. performed the measurements at high magnetic fields.
D.M.M., V.L.M., and J.W. performed the measurements at low magnetic fields.
W.B. performed the theoretical analysis.
The writing of the paper was led by W.B. with input from all authors.


\bibliography{3DTI-sub-nl}


\clearpage
\begin{center}
{\large\bf Supporting Information for\\[-.25em]}
{\LARGE\bf ``Massive and topological surface states\\[-.5em]in tensile strained HgTe''\\}
{}
\vspace{1em}

{\large\sf David M.\ Mahler, Valentin L.\ M\"uller, Cornelius Thienel, Jonas Wiedenmann,\\
Wouter Beugeling, Hartmut Buhmann, and Laurens W.\ Molenkamp}

\end{center}

\section{Modelling of the electrostatics}
The distribution of carriers in a stacked geometry is typically described by the \emph{charge} density $\rho(z)$, where $z$ is the coordinate in the growth direction; it is given in units of $e\,\mathrm{nm}^{-3}$ or $e\,\mathrm{cm}^{-3}$.
The carrier density $n$ (units $\mathrm{nm}^{-2}$ or $\mathrm{cm}^{-2}$) is the integral over the $z$ coordinate,
\begin{equation}
  n = -\frac{1}{e}\int_{z_1}^{z_2}\rho(z) dz,
\end{equation}
where $z_1$ and $z_2$ are the $z$ coordinates of the bottom and top of the sample, and the division by $-e$ is necessary in order to obtain a \emph{particle} density.

The electric field $\vec{E}$ associated to the charge density $\rho(z)$ is found from Maxwell's equation (Gauss' law) $\nabla \cdot \vec{D} = \rho$, involving the electric displacement field $\vec{D}=\epsilon_0\epsilon \vec{E}$, where $\epsilon_0$ is the vacuum permittivity and $\epsilon$ the dielectric constant of the material. The electric potential $V(z)$ satisfies $\vec{E} = -\nabla V$; the corresponding energy potential equals $U(z)=-eV(z)$. We thus find the Poisson equation for $U(z)$,
\begin{equation}
  \ddz\bigl(\epsilon(z)\ddz U(z)\bigr) = \frac{e}{\epsilon_0}\rho(z),
\end{equation}
where we take $\epsilon(z)$ to be a step-wise function that models the dielectric constants of the layers in the layer stack, see Fig.~2 of the main text.

Solution of the Poisson equation is obtained by two-fold integration
\begin{equation}
  U(z) = \int_{z_0}^{z}\mathrm{d}z'
         \frac{e}{\epsilon_0\epsilon(z')}\int_{z'_0}^{z'}\mathrm{d}z''\rho(z''),
\end{equation}
where $z_0$ and $z'_0$ are arbitrary constants. In order to find a unique solution, we need to fix the two integration constants by imposing suitable boundary conditions. For the purpose of this discussion, we set $z_0=z'_0=z_\mathrm{b}$ where $z_\mathrm{b}=-\tfrac{1}{2}d_\mathrm{TI}$ is the $z$-coordinate of the bottom surface of the TI layer. This sets potential and electric field to zero at this location, $U(z_\mathrm{b})=0$ and $\ddz U(z_\mathrm{b})=0$, respectively.

We will make the assumption that all free carriers are in a layer of thickness $d_\mathrm{s}$ under the top surface of the TI at $z=z_\mathrm{t}=\tfrac{1}{2}d_\mathrm{TI}$, i.e., $\rho(z)=0$ for $z<z_\mathrm{t}-d_\mathrm{s}$ or $z>z_\mathrm{t}$. Thus, $U(z)=0$ for $z<z_\mathrm{t}-d_\mathrm{s}$, while for $z>z_\mathrm{t}$, we find that the derivative satisfies
\begin{equation}
  \ddz U(z) = -\frac{e^2}{\epsilon_0\epsilon(z)}n_\mathrm{s},
\end{equation}
where $n_\mathrm{s}$ is the total density. This equation is valid regardless of the density profile (shape of $\rho(z)$). The potential value at the interface is
\begin{equation}\label{eq:ui_ns}
  \Ui \equiv U(z_\mathrm{t})
  =-c \frac{e^2}{\epsilon_0\epsilon_\mathrm{s}}n_\mathrm{s} d_\mathrm{s},
\end{equation}
where $c$ is a coefficient depending on the shape of $\rho(z)$. For a density profile that is uniform in the top surface region, i.e., $\rho(z)=-en_\mathrm{s}/d_\mathrm{s}$ for $z_\mathrm{t}-d_\mathrm{s}<z<z_\mathrm{t}$ and $\rho(z)=0$ elsewhere, as illustrated in Fig.~2a of the main text, we have $c = \tfrac{1}{2}$.
For this choice of $\rho(z)$, the potential $U(z)$ is quadratic in the top surface region.

Since there are no free charges in the cap and insulator layer, the displacement field is constant for $z>z_\mathrm{t}$. Putting the gate at the top surface of the insulator layer, $z=z_\mathrm{g} = \tfrac{1}{2}d_\mathrm{TI}+d_\mathrm{cap}+d_\mathrm{ins}$, we thus find the gate potential to be
\begin{equation}
  U_\mathrm{g} \equiv U(z_\mathrm{g}) = \frac{-e^2}{\epsilon_0}n_\mathrm{s}\left(c \frac{d_\mathrm{s}}{\epsilon_\mathrm{s}} + \frac{d_\mathrm{cap}}{\epsilon_\mathrm{cap}} + \frac{d_\mathrm{ins}}{\epsilon_\mathrm{ins}}\right).
\end{equation}
This potential is related to the gate voltage $\Vg$ as $U_\mathrm{g}=-e\Vg$. Thus, we find the response of the potential to a gate voltage change,
\begin{equation}
  \frac{\Delta \Ui}{\Delta \Vg}
  = \frac{-e\,c\,d_\mathrm{s}/\epsilon_\mathrm{s}}{c\,d_\mathrm{s}/\epsilon_\mathrm{s} + d_\mathrm{cap}/\epsilon_\mathrm{cap} + d_\mathrm{ins}/\epsilon_\mathrm{ins}}.
\end{equation}
and the response of the surface state density (also known as gate action),
\begin{equation}
  \frac{\Delta n_\mathrm{s}}{\Delta \Vg}=\frac{\epsilon_0}{e}\bigl({c\,d_\mathrm{s}/\epsilon_\mathrm{s} + d_\mathrm{cap}/\epsilon_\mathrm{cap} + d_\mathrm{ins}/\epsilon_\mathrm{ins}}\bigr)^{-1}.
\end{equation}
With the parameters for the cap and insulator layer
substituted, we have
\begin{equation}\label{eq:ui_vg}
  \frac{\Delta \Ui}{\Delta \Vg}
  =\frac{-e\,c\,d_\mathrm{s}/\epsilon_\mathrm{s}}{c\,d_\mathrm{s}/\epsilon_\mathrm{s} + 22.91\,\mathrm{nm}}
  =-e\left(1+\frac{\epsilon_\mathrm{s}}{c} \frac{22.91\,\mathrm{nm}}{ d_\mathrm{s}}\right)^{-1}.
\end{equation}
and
\begin{equation}
  \frac{\Delta n_\mathrm{s}}{\Delta \Vg}=\frac{\epsilon_0}{e}\bigl(c\,d_\mathrm{s}/\epsilon_\mathrm{s} + 22.91\,\mathrm{nm}\bigr)^{-1}.
\end{equation}
From Eq.~\ref{eq:ui_vg}, we observe that for a fixed gate voltage, the strength $\Ui$ of the potential in the surface region of the TI increases as the dielectric constant $\epsilon_\mathrm{s}$ is decreased.

For the conjectured surface state properties $\epsilon_\mathrm{s} = 3$ and $d_\mathrm{s} = 8\,\mathrm{nm}$, together with $c=\frac{1}{2}$ for a uniform density profile, we find $\Delta \Ui/\Delta \Vg\approx -1/18.2\,\mathrm{eV}/\mathrm{V} \approx -55 \,\mathrm{meV}/\mathrm{V}$. The gate action is $\Delta n_\mathrm{s}/\Delta \Vg = \SI{2.3e11}{cm^{-2}V^{-1}}$.
If we had chosen the bulk dielectric constant, $\epsilon_\mathrm{s}=\epsilon_\mathrm{HgTe} = 20.8$, the result would be
$\Delta \Ui/\Delta \Vg\approx -1/120\,\mathrm{eV}/\mathrm{V} \approx -8.3 \,\mathrm{meV}/\mathrm{V}$. The signatures of the Volkov-Pankratov states in the experiments indicate that the response $\Delta \Ui/\Delta \Vg$ is stronger than the latter value, so that we can deduce that the surface states are subject to a lower dielectric constant than the  bulk value. This provides an additional confirmation of the observations of Ref.~\cite{Brune2014}.

The Fermi energies $E_\mathrm{F}$ as shown in Fig.~3 of the main text are determined as follows.
From Eq.~\ref{eq:ui_ns}, we find that $\Delta n_\mathrm{s}/\Delta\Ui=\SI{-0.0414e11}{cm^{-2}meV^{-1}}$. For $\Ui=0\,\mathrm{meV}$, we set the density $n_\mathrm{s}=n_{\mathrm{s},0}=\SI{2.5e11}{cm^{-1}}$. The relation between $\Ui$ and the surface state density thus becomes $n_\mathrm{s}(\Ui)=n_{\mathrm{s},0}+(\Delta n_\mathrm{s}/\Delta\Ui)\Ui$. The dispersion (with applied potential strength $\Ui$) yields the integrated density of states as function of energy $n(E)$. The Fermi energy $E_\mathrm{F}$ follows from solving $n(E_\mathrm{F}) = n_\mathrm{s}(\Ui)$ numerically.

\clearpage
\vspace{10mm}
\section{Properties of sample 1}
\label{sec:densitysample1}
\begin{table}[hb]
  \begin{tabular}{S[table-format=2.2]|S[table-format=3.2]|S[table-format=3.1]}
    {$\Vg$} & {$n_\text{tot}$} & {$\mu_\text{tot}$}\\
    {(V)} & \multicolumn{1}{c|}{($10^{11}\,\mathrm{cm}^{-2}$)} & {($10^3\,\mathrm{cm}^2\,\mathrm{V}^{-1}\mathrm{s}^{-1}$)}\\
    \hline
    2.5 &  7.96 & 89.6 \\
    1.5 &  5.56 & 75.9 \\
    0.0 &  2.44 & 115.0 \\
   -2.0 & -4.83 & 45.2 \\
   -3.0 & -5.95 & 34.9 \\
   -4.0 & -7.60 & 27.1 \\
   -5.0 & -9.67 & 24.2 \\
   -6.0 & -12.2 & 26.6 \\
   -7.0 & -14.6 & 31.3 \\
   -8.0 & -16.3 & 32.4 \\
   -9.0 & -18.6 & 31.5 
  \end{tabular}
  \caption{Total density $n_\text{tot}=n_\text{e}-n_\text{h}$ as function of gate voltage $\Vg$ for sample~1. We also provide the total mobility $\mu_\text{tot}$. The densities  have been obtained from a linear fit of $R_{xy}$ in the classical Hall regime ($B\lesssim\SI{1}{T}$). The average mobilities have been estimated from $\mu_\text{tot}=1/(e n_\text{tot} \rho_{xx})$, where $\rho_{xx}=R_{xx}/A$ is the longitudinal resistivity at $B=\SI{0}{T}$ ($A$ is the aspect ratio).}
  \label{table:densitysample1}
\end{table}

\section{Properties of sample 2}
\label{sec:densitysample2}
\begin{table}[hb]
  \begin{tabular}{S[table-format=2.1]|S[table-format=1.2]S[table-format=2.1]S[table-format=3.1]|S[table-format=3]S[table-format=2]}
    {$\Vg$} & {$n_\text{e}$} & {$n_\text{h}$} & {$n_\text{tot}$} & {$\mu_\text{e}$} & {$\mu_\text{h}$}\\
    {(V)} & \multicolumn{3}{c|}{($10^{11}\,\mathrm{cm}^{-2}$)} & \multicolumn{2}{c}{($10^3\,\mathrm{cm}^2\,\mathrm{V}^{-1}\mathrm{s}^{-1}$)}\\
    \hline
    -2.0 & 0.60 & 11.1 & -10.5 & 266 & 15 \\
    -4.0 & 0.58 & 15.7 & -15.2 & 535 & 21 \\
    -6.0 & 0.58 & 20.1 & -19.4 & 562 & 20 \\
    -8.0 & 0.57 & 24.5 & -23.9 & 556 & 18
  \end{tabular}
  \caption{Electron and hole density, $n_\text{e}$ and $n_\text{h}$ respectively, and total density $n_\text{tot}=n_\text{e}-n_\text{h}$ as function of gate voltage $\Vg$ for sample~2. We also provide the mobilities for electrons and holes, $\mu_\text{e}$ and $\mu_\text{h}$, respectively. The total density $n$ has been obtained from a linear fit of $R_{xy}$ for $B=\SI{0.3}{T}$ to $\SI{1}{T}$. The electron density has been obtained from the $1/B$ periodicity of $R_{xx}$. The mobilities were obtained from Eq.~(2) of the main text together with the linear fits for low magnetic fields shown in Fig.~5 of the main text and $R_{xx}$ at $B=\SI{0}{T}$. }
  \label{table:densitysample2}
\end{table}

\end{document}